\documentclass[aps,prb,twocolumn,superscriptaddress,showpacs]{revtex4}
\usepackage{amsmath}
\usepackage{graphicx}

\begin{document}

\title{Pairing symmetry signatures of $T_{1}$ in superconducting ferromagnets}

\author{Hari P. Dahal}
\affiliation{Department of Physics, Boston College, Chestnut Hill, MA, 02467}

\author{Jason Jackiewicz}
\affiliation{Department of Physics, Boston College, Chestnut Hill,
MA, 02467}

\author{Kevin S. Bedell}
\affiliation{Department of Physics, Boston College, Chestnut Hill, MA, 02467}

\date{\today}

\begin{abstract}
We study the nuclear relaxation rate $1/T_{1}$ as a function of temperature for a
superconducting-ferromagnetic coexistent system using a p-wave triplet model for the superconducting pairing symmetry.  This calculation is contrasted with a singlet s-wave one done previously, and we see for the s-wave case that there is a Hebel-Slichter peak, albeit reduced due to the magnetization, and no peak for the p-wave case.  We then compare these results to a nuclear relaxation rate experiment on $UGe_{2}$ to determine the possible pairing symmetry signatures in that material.  It is seen that the experimental data is inconclusive to rule out the possibility of s-wave pairing in $UGe_{2}$.
\end{abstract}


\maketitle

\begin {figure}[t!]
\includegraphics[width= 8.0 cm,height = 6.0cm]{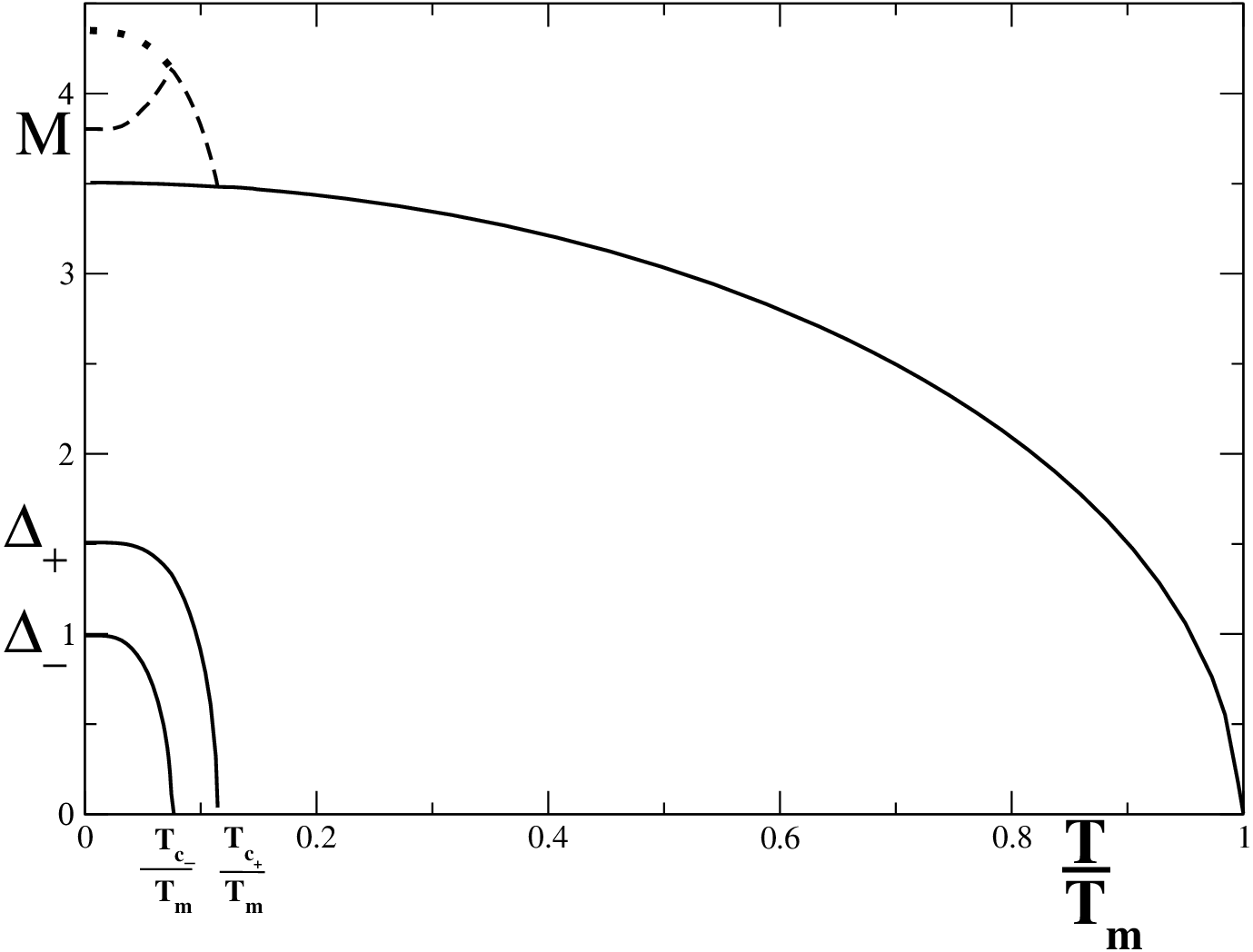}
\caption{Order parameters in this model as a function of $T$. $M$, $\Delta_{+}$,
and $\Delta_{-}$ are labeled accordingly. For
M, the solid curve refers to the pure ferromagnetic state, $\Delta_{+} = \Delta_{-} = 0$,
the dashed line refers to $M$ when $\Delta_{+} > \Delta_{-} \neq 0$, and
the dotted line when $\Delta_{-} = 0$ and $\Delta_{+} \neq
0$. The y-axis is a unitless measure and the x-axis is scaled by the Curie temperature ($T_{m}$).}
 \label{odparameter}
\end{figure}

In the BCS theory of superconductivity\cite{bcs1957}, the
conduction electrons in a metal cannot be both ferromagnetically
ordered and superconducting. Superconductors may expel small magnetic fields
passing through them, but strong magnetic fields destroy the
superconductivity (SC). Even small amounts of magnetic impurities
are usually enough to eliminate SC. Much work has been done both
theoretically and experimentally to understand this interplay, and
to explore the possibility of coexistence between these two
ordered states.\\
\indent The archetypal material that does display this coexistence is $UGe_{2}$, and it is generally believed that the superconductivity is governed by triplet pairing due to the presence of the internal ferromagnetism\cite{huxley2001}.  However, there have been some convincing calculations that instead consider s-wave superconductivity to be favored in the presence of this magnetization.  For instance, Abrikosov\cite{abrikosov2001} and Suhl\cite{suhl2001} demonstrate that $UGe_{2}$ should have s-wave Cooper pairing which is mediated by localized ferromagnetic spins. Also, Blagoev \textit{et al.}\cite{blagoev1998} studied a
weak ferromagnetic Fermi liquid and showed that s-wave
superconductivity is possible and subsequently energetically favored in the ferromagnetic
state.\\
\indent The contributions of all of these theories and calculations have naturally led to a healthy debate on this specific property of this fascinating material, and thus the nature of the pairing is still an open question\cite{flouquet2005}. \\
\indent Motivated by this debate and the experimental data that should hold the answer, the present authors have recently studied various properties of a mean-field model of coexistent itinerant ferromagnetism and s-wave superconductivity\cite{karchev2001}. The electronic specific heat was calculated\cite{jackiewicz2003} and quantitatively fit the experimental data quite well, as well as several other properties, including the nuclear relaxation rate\cite{dahal2005}.\\
\indent In this paper we show the results of a nuclear
relaxation rate $1/T_{1}$ calculation from a different starting model\cite{andriy2004}
in which there is p-wave pairing in the superconducting channel.  This result is then compared to the s-wave calculation and the experimental data to determine a signature of the much disputed pairing symmetry.  We find that there is a small signature around $T_{c}$ in the observed data of a small Hebel-Slichter peak that resembles the s-wave curve, but that this comparison is inconclusive to verify either type of pairing.\\
\indent A triplet state can be described by the order parameter
$\hat{\Delta}_{\alpha\beta}(\textbf{k})= \langle
c_{\textbf{k},\alpha}c_{\textbf{k},\beta}\rangle = [\textit{i}
(\textbf{d}(\textbf{k}).\sigma)\sigma_{y}]_{\alpha\beta}$, where $
\sigma = \hat{x}\sigma_{x}+\hat{y}\sigma_{y}+\hat{z}\sigma_{z}$
denotes the usual Pauli matrices, and the basis of symmetric
matrices $ \textit{i}\sigma\sigma_{y}$ is used to represent odd
angular momentum pairing. The three-dimensional complex vector
$\textbf{d}(\textbf{k})$ fully characterizes the triplet pairing
state. For simplicity, the easy axis of magnetization is assumed to
be in the z-direction. Because of the pair-breaking effect of the
strong exchange field, only the Cooper pairs with equal
spin states will survive. In the case of equal spin pairing we can write
vector $\textbf{d}$ in the form $\textbf{d} = (d_{x},d_{y},0)$.
Denoting $\Delta_{\pm} \equiv d_{x} \pm d_{y}$, the
superconducting order parameter (SCOP) becomes\cite{andriy2004}
\begin{equation}
\hat{\Delta}_{\alpha\beta}(\textbf{k})= \left(%
\begin{array}{cc}
  -\triangle_{-}(\textbf{k}) & 0 \\
  0 & \triangle_{+}(\textbf{k}) \\
\end{array}%
\right).
\end{equation}
\indent An attractive pair-forming interaction $V$ is assumed, and the coexistent Hamiltonian is then
\begin{eqnarray}
\nonumber
 H_{FM+SC}= \sum_{\textbf{k},\alpha}\epsilon_{\textbf{k}}
c_{\textbf{k},\alpha}^{\dag}c_{\textbf{k},\alpha} - I\int
d\textbf{r} s(\textbf{r})\cdot s(\textbf{r}) \\
+\frac{1}{2}\sum_{\textbf{k},\textbf{k}^{'}}V_{\alpha\beta,\lambda\mu}(\textbf{k},\textbf{k}^{'})
c_{\textbf{-k},\alpha}^{\dag}c_{\textbf{k},\beta}^{\dag}c_{\textbf{k}^{'},\lambda}c_{\textbf{-k}^{'},\mu}.
\end{eqnarray}
\indent The equations for the SCOP have
the usual BCS-like form:
\begin{equation}\label{supconodparamm}
 \Delta_{-}(\textbf{k})=
\frac{-1}{V}\sum_{\textbf{k}^{'}}V(\textbf{k},\textbf{k}^{'})\frac{1-2f(E_{-}({\textbf{k}}^{'}))}{2
E{-}({\textbf{k}}^{'})}\Delta_{-}({\textbf{k}}^{'})\\
\end{equation}
\begin{equation}\label{supconodparamp}
\Delta_{+}(\textbf{k})=
\frac{-1}{V}\sum_{{\textbf{k}}^{'}}V(\textbf{k},{\textbf{k}}^{'})\frac{1-2f(E_{+}({\textbf{k}}^{'}))}{2
E{+}({\textbf{k}}^{'})}\Delta_{+}({\textbf{k}}^{'}),
\end{equation}
where $f(E)$ is the Fermi-Dirac distribution function, and
$\Delta_{+}$ and $\Delta_{-}$ are the SCOP for the up-up and the
down-down spin pairing states respectively. The equation for the magnetic
order parameter (MOP) is
\begin{equation}\label{magodparam}
 M =
\frac{1}{V}\sum_{\textbf{k}}\left\{\frac{\epsilon_{\textbf{k}}^{\uparrow}[1-2f(E_{-})]}{2E_{-}(\textbf{k})
} -
\frac{\epsilon_{\textbf{k}}^{\downarrow}[1-2f(E_{+})]}{2E_{+}(\textbf{k})}\right\},
\end{equation}
where $\epsilon_{\textbf{k}^{\uparrow,\downarrow}}\equiv
\epsilon_{\textbf{k}} \pm \frac{IM}{2}$.
Eqs.(\ref{supconodparamm},\ref{supconodparamp},\ref{magodparam})
are coupled with each other via the quasiparticle spectrum
$E_{\pm}(\textbf{k})= \sqrt{(\epsilon_{\textbf{k}}\mp
\frac{IM}{2})^{2} + |\Delta_{\pm}(\textbf{k})|^{2}}$.\\
\indent In order to
illustrate the interplay between the MOP and the SCOP, we solve
eqs.(\ref{supconodparamm},\ref{supconodparamp},\ref{magodparam})
self-consistently for the simplest case of a spherical Fermi
surface, assuming that the SC pairing strength is in the p-channel,
$V(\textbf{k},{\textbf{k}}^{'})=
g_{l=1}(k,k')\sum_{m=-1}^{1}Y_{1m}(\hat{\textbf{k}})Y_{1m}^{\ast}(\hat{{\textbf{k}}^{'}})$,
and has the BCS-like form, i.e., $g_{1}(k,k')$ vanishes everywhere
except the narrow region near the Fermi surface and does not
depend on the exchange interaction $I$. We have used
$\Delta_{\pm}(\textbf{k})= \Delta_{0\pm}Y_{1}^{\pm1}$. The
SCOP will have point nodes at the poles of the Fermi surface.\\
\indent The order parameters, $M$ and $\Delta$, have dependencies such as
$M = M(g,I,T)$ and $\Delta = \Delta(g,I,T)$ at $T\neq 0$. After solving these three coupled integral equations numerically,  we study the
characteristic behavior of the order parameters.\\
\indent The variation of the transition temperature of the ferromagnetic
and the superconducting transitions as a function of $I$ has been
studied previously\cite{andriy2004}, and here we mention the results. For $I\leq 1.0$, which corresponds
to the paramagnetic phase $M=0$, the superconducting
transition temperature for the up-up, $T_{c_{+}}$, and down-down,
$T_{c_{-}}$, spin paring is equal, implying that $\Delta_{+} =
\Delta_{-} $. This corresponds to the planer (unitary) phase of
the triplet pairing. But for $I>1.0$, there is a finite magnetization
and the transition temperatures $T_{c_{+}}$ and $T_{c_{-}}$ will
not be equal, implying that $\Delta_{+} \neq \Delta_{-} $.
Upon further increase of $I$, and the subsequent increase of $M$,  $T_{c_{+}}$
increases and $T_{c_{-}}$ decreases to zero. Therefore, the
finite magnetization induces the non-unitary type of
superconducting triplet state.\\
\indent Here, we study the temperature dependence of $M$, $\Delta_{+}$, and
$\Delta_{-}$ for fixed values of $I$ and $g$. For $I\leq1.0$,
in the paramagnetic phase, the variation of $\Delta_{+}$ and
$\Delta_{-}$ is identical to that of the mean field
result $\Delta \propto (T_{c}-T)^{\frac{1}{2}}$ as in the BCS case.
Similarly, the temperature variation of $M$ in the absence of
superconductivity, i.e., $\Delta_{+}=\Delta_{-}=0$, follows the
trend as is predicted by mean field theory (solid line in
Fig.(\ref{odparameter})). We show the
variation of $M$, $\Delta_{+}$, and $\Delta_{-}$ as a function
of $T$ for the chosen values of $I=1.018$ and $g=0.95$ (coexistence
of magnetism and both equal spin pairing triplet state superconductivity) in Fig.(\ref{odparameter}) . The
temperature axis has been normalized with the ferromagnetic
transition (Curie) temperature $T_{m}$ to show the relative magnitudes of
the corresponding transition temperatures. We can tune values of
$I$ and $g$ so that the transition temperatures will be
comparable.\\
\indent We see in Fig.(\ref{odparameter}) that the SCOP decrease
with temperature. The MOP decreases as well for
$T \geq T_{c_{+}}$, but for $T \leq T_{c_{+}}$ we see some new
features. If
 $\Delta_{+} > \Delta_{-} \neq 0$, for $T \leq T_{c_{+}}$, $M$ (dashed line) is
always bigger than its value in the pure ferromagnetic state when $\Delta_{+} = \Delta_{-} = 0$ (solid line). When there is a finite $\Delta_{+}$ and $\Delta_{-}$, $M$ increases
with the increase in $T$ from $0$ to $T_{c_{-}}$, and then decreases sharply
in the range of temperature $T_{c_{-}} < T < T_{c_{+}}$(dashed
line). The increase in $M$ for $0 < T \leq T_{c_{-}}$ is due to
the decreasing $\Delta_{-}$ as the temperature increases.
On the other hand, the sharp decrease in $M$ for $T_{c_{-}} \leq T
\leq T_{c_{+}}$ is not only due to the temperature, but
more strongly to the decrease of $\Delta_{+}$. Again we see that for $0 < T \leq T_{c_{-}}$, $M$
is bigger if $\Delta_{-} = 0$ (dotted line) than when $\Delta_{-}
\neq 0$ (dashed line). These results are in agreement with
the general intuition that the magnetization has a cooperative
relation with the up spin pairing state, and a competitive relation with
the down spin pairing.
Our ultimate goal is to study the variation of the relaxation rate $1/T_{1}$
with temperature for this p-wave state, so next we derive the equation for the
density of states as a function of excitation
energy.\\
\indent To derive the expressions for the density of states we use the
usual relation,
$N_{s}^{\sigma}(E)= \frac{1}{({2\pi})^{3}}\int{d\bar{p}\delta(E-E_{p}^{\sigma})}$, where $\sigma$ refers to the up spin or down spin fermions. Solving in the standard way using the property of the $\delta-$function, we get,
\begin{equation}
 N_{s}^{\uparrow,\downarrow}(E) = \frac{1}{8\pi^{2}}
\int_{0}^{\pi}d\theta\sin\theta\left(A^{+}+A^{-}\right),
\end{equation}
where
\begin{eqnarray}
\nonumber
A^{+}& = & \frac{{\sqrt{{p_{F}^{2} + 2m^{*}\left(\frac{\pm
 IM}{2} + \sqrt{E^{2}-\Delta_{\pm}^{2}}\right)}}
 }}{\left|\frac{\sqrt{E^{2}-\Delta_{\pm}^{2}}}{2m^{*}E}\right|},\\
A^{-} & = &  \frac{{\sqrt{{p_{F}^{2} + 2m^{*}\left(\frac{\pm
IM}{2}-\sqrt{E^{2}-\Delta_{\pm}^{2}}\right)}}
 }}{\left|\frac{\sqrt{(E^{2}-\Delta_{\pm}^{2})}}{2m^{*}E}\right|}.
\end{eqnarray}
The $+$ sign
is for the $\uparrow$ spin fermions and the $-$ sign is for
$\downarrow$ spin fermions on $\Delta_{\pm}$, respectively. These two expressions converge to the density of states of the
normal state at the Fermi level, $N(0) =\frac{m^{*}p_{f}}{\pi^{2}}$,
in the limit of all OP equal to zero. For $M=0$, the expressions
result in the density of states of the equal spin pairing
triplet state superconductor, $N(E)/N(0) = E/2\int d\theta\sin\theta/\sqrt{(E^{2} -
\Delta^{2})}$. The important effect of the internal magnetization $M$ is to modify the value of the density of
states.  For increasing $M$, we observe an increase in $N_{s}^{\uparrow}(E)$ and a decrease in $N_{s}^{\downarrow}(E)$.\\
\begin{figure}[t]
\includegraphics[width=7.7 cm,height=6.cm]{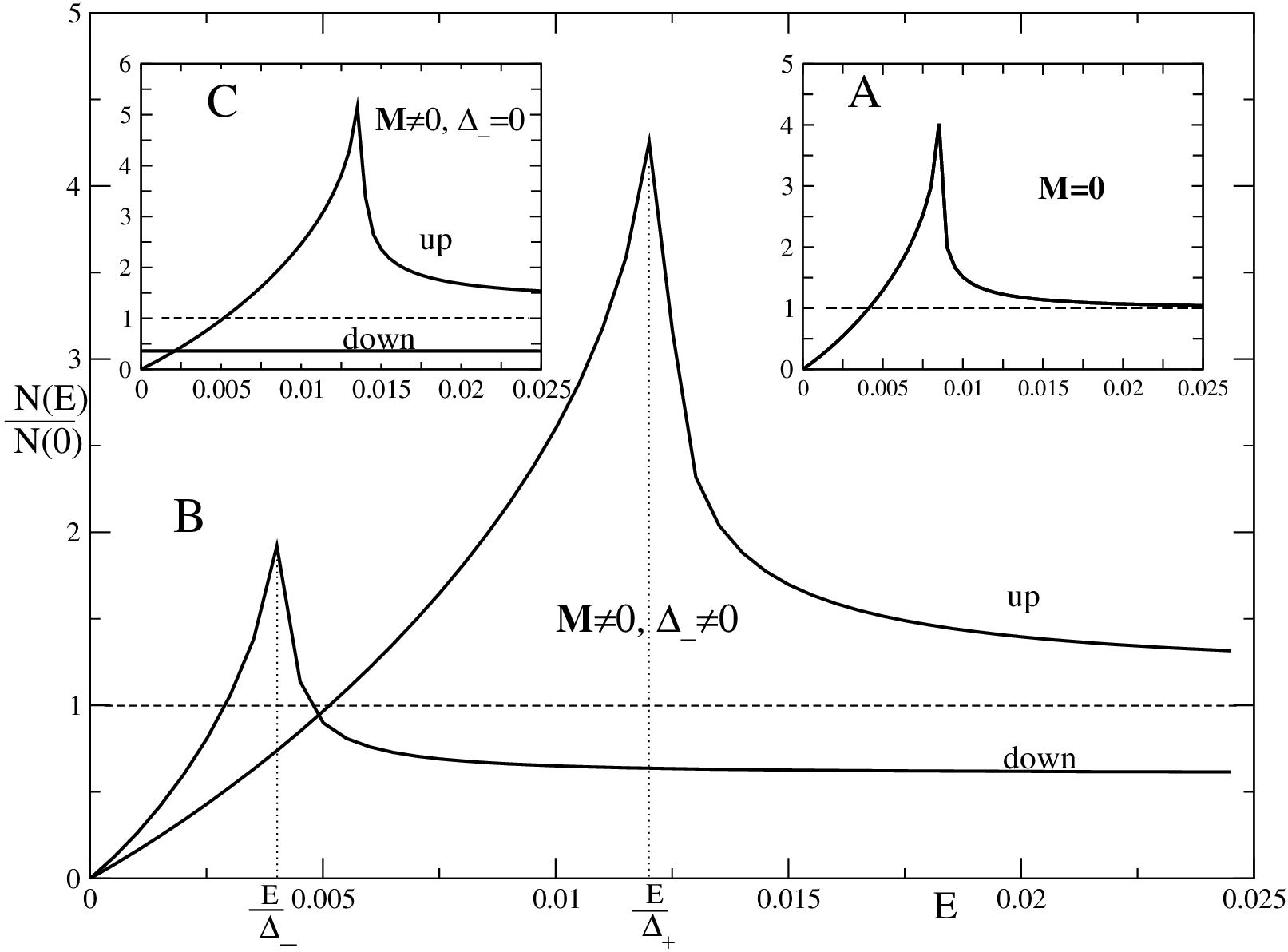}
 \caption{Density of states as a function of the excitation energy for each spin species.
The dotted line corresponds to the normal state at the Fermi level.
Inset A corresponds
to $M=0$, B to $M \neq 0$ for which down spin pairing is finite
and C to $M \neq 0$ for which down spin pairing is completely
suppressed. Note that the density of states of the up and the down spin
fermions are not identical when $M \neq 0$.}
\label{densityofstates}
\end{figure}
\indent We plot the $T=0$ density of states (DOS) scaled to the normal state value for positive excitation energies in Fig.(\ref{densityofstates}).
The inset A represents the density of
states when $M = 0$, which is just a superconducting DOS and both spin components fall on the same line.  The divergence
in the density of states is at $E=\Delta_{+}=\Delta_{-}$. The
main body B of Fig.(\ref{densityofstates}) shows the effect of the magnetization on the
density of states when the magnitude of $M$ is such that $\Delta_{-}
\neq 0$. We see that due to the finite magnetization, the density of
states curve now splits for the up and the down spin fermions. The
diverging density of states at the lower energy is at
$E=\Delta_{-}$ and that at the higher value is at
$E=\Delta_{+}$. For the value of $I$ at which $M$ is large enough to completely suppress the down spin
pairing, the density of states for the down spin fermion
will be constant as is shown in inset C.
These modifications in the density of
states due to the presence of the internal magnetization will manifest
themselves in the nuclear relaxation rate of
the system.\\
\indent In general $1/T_{1}$ in the superconducting state is
related to the (DOS) in the superconducting
state, $N_{s}(E)$ as follows \cite{tinkham}:

\begin{equation}
\frac{1}{T_{1}} \propto 2 \int_{0}^{\infty} (N_{s}(E))^2
f(E)(1-f(E))dE,
 \end{equation}
 where $f(E)$ is the Fermi-distribution function.
 In the presence of $M$ the functions will not be the same
 for the
 up and the down spin fermions as mentioned above, so we split the integral for different spin species. Hence,

 \begin{equation}
 \begin{split}
\frac{1}{T_{1}} \propto \int_{0}^{\infty}
(N_{s}^{\downarrow}(E))^2f(E_{-})(1-f(E_{-}))dE \\
+ \int_{0}^{\infty}
(N_{s}^{\uparrow}(E))^2f(E_{+})(1-f(E_{+}))dE.
\end{split}
 \end{equation}

\begin{figure}[b]
\includegraphics[width=7.5 cm,height=6.2cm]{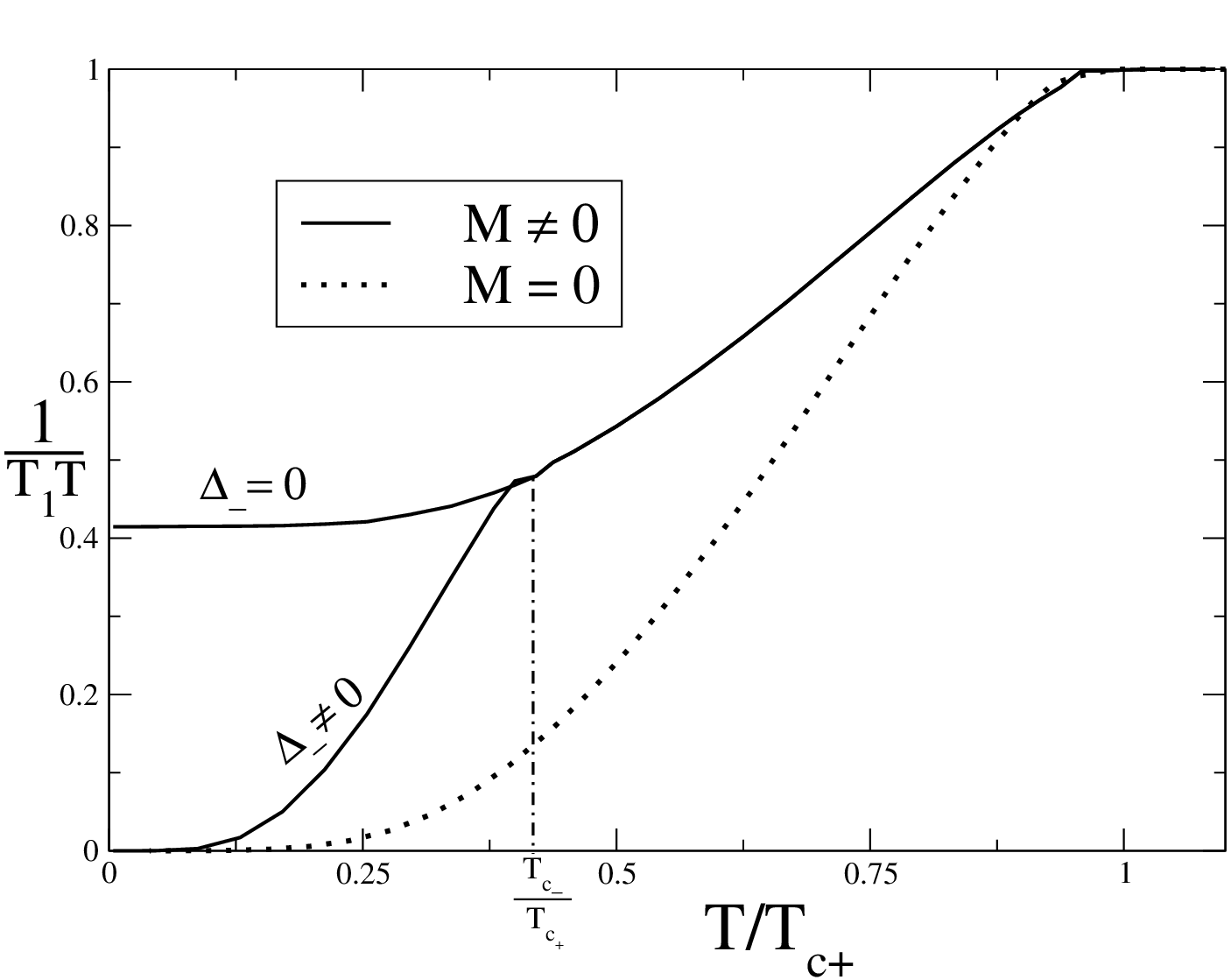}
\caption{$1/T_{1}T$ vs. normalized temperature with respect to the
transition temperature of the up spin Cooper pairs, $T_{c+}$.  The dotted line is the p-wave superconducting case, with no magnetization.  The top solid line is for the case when there is no pairing of down spins, where the only transition is at $T/T_{c+}=1$ when the system becomes an up spin paired superconductor, and saturates at $T=0$.   The bottom solid curve shows a second transition when the down spin Cooper pairs are formed and another gap $\Delta_{-}$ opens up, and is fully suppressed to zero. The values of
$I$ and $g$ are chosen so that $T_{c+}/T_{c-}=2.5$.}
\label{T1T}
\end{figure}
First, we choose the interaction parameters $I$ and $g$ such that
 there is both the up-up spin and the down-down spin pairing
 (the corresponding transition temperatures are not
 equal) and study the variation of $1/T_{1}T$
 with respect to temperature. The results have
 been presented in Fig.(\ref{T1T}), where the temperature has been normalized with the
 transition temperature for the up-up spin pairing.  We choose $I$ and $g$ such
 that $T_{c+}/T_{c-}\sim 2.5$. In Fig.(\ref{T1T}) we see that $1/T_{1}T$
 decreases for $T \leq T_{c+}$ . This decrement characterizes the onset of
 the spin triplet pairing. With the decrease in temperature the rate continues to decrease, but not as strongly as it would for the $M=0$ case, as shown by the dotted line.  The dotted line is just the p-wave superconducting case, and when the gap opens up, the rate is suppressed quickly.\\
 \indent On the other hand, when we have a finite magnetization and only superconducting pairing of the up spins, we refer to the top solid curve of Fig.(\ref{T1T}).  In this case, the transition at $T=T_{c+}$ results in a decrease of the relaxation rate due to the presence of the gap $\Delta_{+}$, but the rate does not decrease as sharply as the unpolarized case because there is a finite magnetization.  Furthermore, there is a normal fermionic component in this case since the down spins have not paired (due to our choice of the interaction parameters), and thus this finite density at the Fermi level allows for a nonzero relaxation rate at zero temperature.  This follows from the fact that the entire Fermi surface is not gapped.\\
 \indent Finally, the bottom solid curve of Fig.(\ref{T1T}) is calculated when we allow for the down spins to pair.  This results in a second superconducting transition at $T_{c-}$ and the opening up of a second gap $\Delta_{-}$.  The relaxation rate goes to zero when all the states are frozen out at zero temperature as expected. Note that there is no Hebel-Slichter peak for the p-wave case.

\begin{figure}
\includegraphics[width=8.0 cm,height=6.cm]{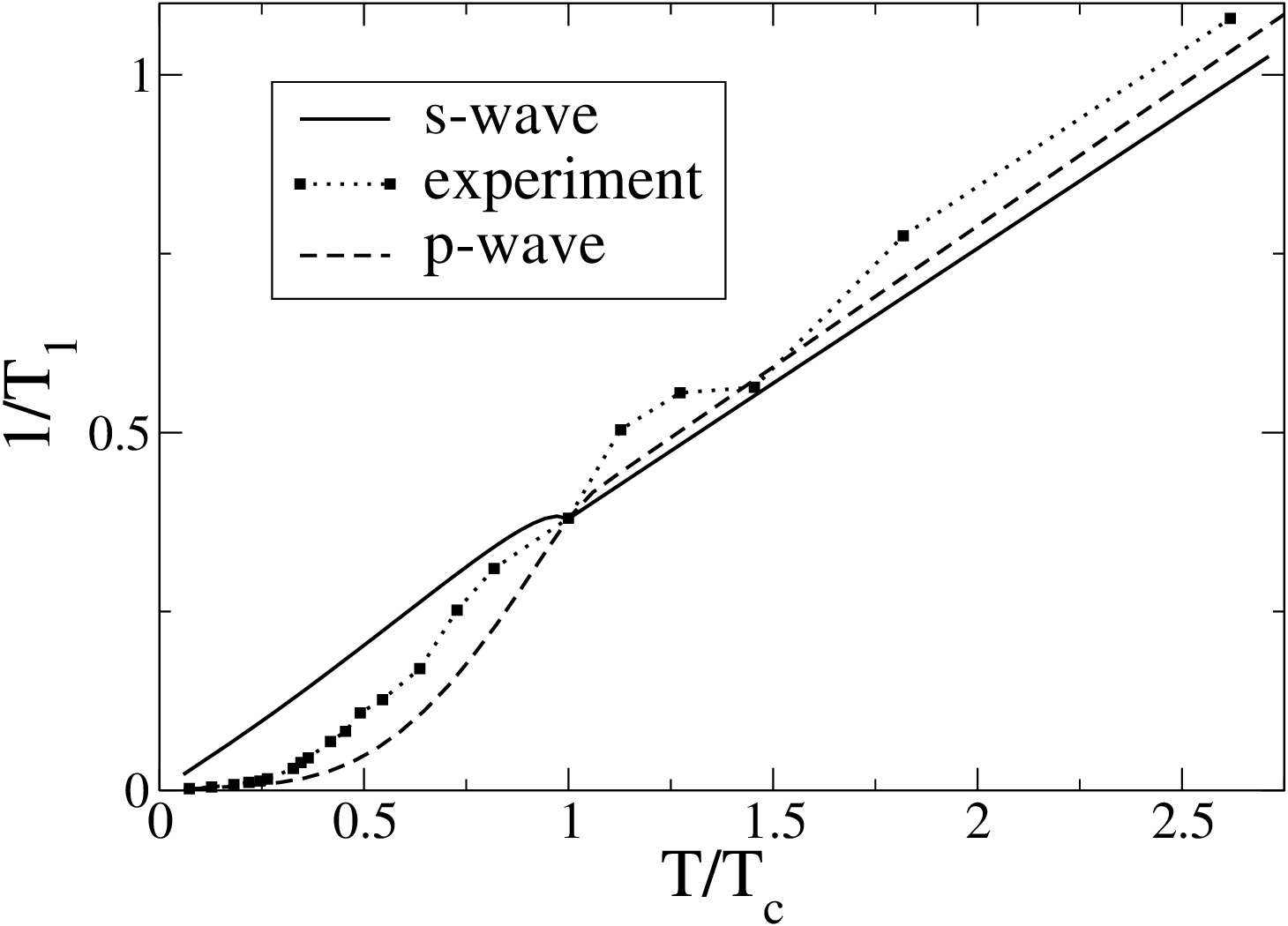}
\caption{Comparison of theoretical and experimental data. The temperature axis is scaled by the SC transition temperature. The experimental data\cite{kotegawa2004} was taken at $13kbar$: at this pressure there is an observed superconducting transition at $T\sim0.5K$}
\label{exp}
\end{figure}

We now compare these results with the s-wave singlet nuclear relaxation rate and the experimental data.  We have received data from the Kotegawa group\cite{kotegawa2004}on $UGe_{2}$.  The calculation of the singlet has been carried out in a prior publication\cite{dahal2005}, where the details can be found.  In Fig.(\ref{exp}) we calculate $1/T_{1}$ versus temperature and show all three curves. The experimental data was taken at $13kbar$.  We emphasize that at this pressure and temperatures, there is coexistence of superconductivity and magnetism.  Also, the theoretical triplet curve is for the case of one single transition at $T=T_{c}$ corresponding to the up spins forming Cooper pairs.  The down spins do not pair in this case and behave basically as normal fermions.\\
\indent Several features are seen in Fig.(\ref{exp}).  First of all, the data is not very consistent near the critical temperature, due to the difficulties in the measurement.  However, a small peak below $T_{c}$ is noticeable, which follows the trend of the s-wave case of the solid line. This very well could be the reduced Hebel-Slichter peak that was predicted due to the presence of the ferromagnetism which produces a finite density of states at the Fermi level due to the existence of gapless fermions\cite{dahal2005}.\\
\indent This p-wave curve is calculated without down spin pairing, and referring to Fig.(\ref{T1T}) where  $1/T_{1}T$ is constant at small temperatures, $1/T_{1}$ goes to zero with $T$ as expected.  The low temperature behavior of the data seems to agree well within the triplet scenario.  We do not observe in the experimental data a second transition and a second gap opening up from the pairing of the down spins.  From our theoretical analysis, we would expect this to happen at a temperature of around $T=0.018K$, however the experiments have only been carried out down to $T=0.04K$.\\
\indent In conclusion, we have studied a p-wave model of superconducting pairing in the background of itinerant ferromagnetism and calculated the nuclear relaxation rate $1/T_{1}$.  Since there is still much debate as to the type of pairing in these systems, we feel a direct comparison between what is expected for different types of pairing and experimental data is the most important goal.  We see that around the critical temperature, the s-wave model agrees well with the experiment, as it does in the specific heat\cite{jackiewicz2003}, but at lower temperatures there are discrepancies with the theory.  The model does make clear that the magnetization suppresses strongly the superconducting signatures, which is certainly being observed experimentally, and a closer look at other electronic properties in conjunction with careful experiments is what is need to determine the nature of the superconducting pairing symmetry.\\
\indent We thank the Kotegawa group for
providing an early draft of their paper\cite{kotegawa2004} on $UGe_{2}$. This work was done with
the support of DOE/DEFG0297ER45636.
\bibliographystyle{apsrev}
\bibliography{references}

\begin{thebibliography}{12}
\expandafter\ifx\csname natexlab\endcsname\relax\def\natexlab#1{#1}\fi
\expandafter\ifx\csname bibnamefont\endcsname\relax
  \def\bibnamefont#1{#1}\fi
\expandafter\ifx\csname bibfnamefont\endcsname\relax
  \def\bibfnamefont#1{#1}\fi
\expandafter\ifx\csname citenamefont\endcsname\relax
  \def\citenamefont#1{#1}\fi
\expandafter\ifx\csname url\endcsname\relax
  \def\url#1{\texttt{#1}}\fi
\expandafter\ifx\csname urlprefix\endcsname\relax\def\urlprefix{URL }\fi
\providecommand{\bibinfo}[2]{#2}
\providecommand{\eprint}[2][]{\url{#2}}

\bibitem[{\citenamefont{Bardeen et~al.}(1957)\citenamefont{Bardeen, Cooper, and
  Schrieffer}}]{bcs1957}
\bibinfo{author}{\bibfnamefont{J.}~\bibnamefont{Bardeen}},
  \bibinfo{author}{\bibfnamefont{L.~N.} \bibnamefont{Cooper}},
  \bibnamefont{and} \bibinfo{author}{\bibfnamefont{J.~R.}
  \bibnamefont{Schrieffer}}, \bibinfo{journal}{Phys. Rev.}
  \textbf{\bibinfo{volume}{108}}, \bibinfo{pages}{1175} (\bibinfo{year}{1957}).

\bibitem[{\citenamefont{Huxley et~al.}(2001)\citenamefont{Huxley, Sheikin,
  Ressouche, Kernavanois, Braithwaite, Calemczuk, and Flouquet}}]{huxley2001}
\bibinfo{author}{\bibfnamefont{A.}~\bibnamefont{Huxley}},
  \bibinfo{author}{\bibfnamefont{I.}~\bibnamefont{Sheikin}},
  \bibinfo{author}{\bibfnamefont{E.}~\bibnamefont{Ressouche}},
  \bibinfo{author}{\bibfnamefont{N.}~\bibnamefont{Kernavanois}},
  \bibinfo{author}{\bibfnamefont{D.}~\bibnamefont{Braithwaite}},
  \bibinfo{author}{\bibfnamefont{R.}~\bibnamefont{Calemczuk}},
  \bibnamefont{and} \bibinfo{author}{\bibfnamefont{J.}~\bibnamefont{Flouquet}},
  \bibinfo{journal}{Phys. Rev. B} \textbf{\bibinfo{volume}{63}},
  \bibinfo{pages}{144519} (\bibinfo{year}{2001}).

\bibitem[{\citenamefont{Abrikosov}(2001)}]{abrikosov2001}
\bibinfo{author}{\bibfnamefont{A.~A.} \bibnamefont{Abrikosov}},
  \bibinfo{journal}{J. Phys. Cond. Matt.} \textbf{\bibinfo{volume}{13}},
  \bibinfo{pages}{L943} (\bibinfo{year}{2001}).

\bibitem[{\citenamefont{Suhl}(2001)}]{suhl2001}
\bibinfo{author}{\bibfnamefont{H.}~\bibnamefont{Suhl}}, \bibinfo{journal}{Phys.
  Rev. Lett.} \textbf{\bibinfo{volume}{87}}, \bibinfo{pages}{167007}
  (\bibinfo{year}{2001}).

\bibitem[{\citenamefont{Blagoev et~al.}(1998)\citenamefont{Blagoev,
  Engelbrecht, and Bedell}}]{blagoev1998}
\bibinfo{author}{\bibfnamefont{K.~B.} \bibnamefont{Blagoev}},
  \bibinfo{author}{\bibfnamefont{J.~R.} \bibnamefont{Engelbrecht}},
  \bibnamefont{and} \bibinfo{author}{\bibfnamefont{K.~S.}
  \bibnamefont{Bedell}}, \bibinfo{journal}{Phil. Mag.}
  \textbf{\bibinfo{volume}{83}}, \bibinfo{pages}{3247} (\bibinfo{year}{1998}).

\bibitem[{\citenamefont{Flouquet et~al.}(2005)\citenamefont{Flouquet, Knebel,
  Braithwaite, Aoki, Brison, Hardy, Huxley, Raymond, Salce, and
  Sheikin}}]{flouquet2005}
\bibinfo{author}{\bibfnamefont{J.}~\bibnamefont{Flouquet}},
  \bibinfo{author}{\bibfnamefont{G.}~\bibnamefont{Knebel}},
  \bibinfo{author}{\bibfnamefont{D.}~\bibnamefont{Braithwaite}},
  \bibinfo{author}{\bibfnamefont{D.}~\bibnamefont{Aoki}},
  \bibinfo{author}{\bibfnamefont{J.}~\bibnamefont{Brison}},
  \bibinfo{author}{\bibfnamefont{F.}~\bibnamefont{Hardy}},
  \bibinfo{author}{\bibfnamefont{A.}~\bibnamefont{Huxley}},
  \bibinfo{author}{\bibfnamefont{S.}~\bibnamefont{Raymond}},
  \bibinfo{author}{\bibfnamefont{B.}~\bibnamefont{Salce}}, \bibnamefont{and}
  \bibinfo{author}{\bibfnamefont{I.}~\bibnamefont{Sheikin}},
  \bibinfo{journal}{cond-mat/0505713}  (\bibinfo{year}{2005}).

\bibitem[{\citenamefont{Karchev et~al.}(2001)\citenamefont{Karchev, Blagoev,
  Bedell, and Littlewood}}]{karchev2001}
\bibinfo{author}{\bibfnamefont{N.~I.} \bibnamefont{Karchev}},
  \bibinfo{author}{\bibfnamefont{K.~B.} \bibnamefont{Blagoev}},
  \bibinfo{author}{\bibfnamefont{K.~S.} \bibnamefont{Bedell}},
  \bibnamefont{and} \bibinfo{author}{\bibfnamefont{P.~B.}
  \bibnamefont{Littlewood}}, \bibinfo{journal}{Phys. Rev. Lett.}
  \textbf{\bibinfo{volume}{86}}, \bibinfo{pages}{846} (\bibinfo{year}{2001}).

\bibitem[{\citenamefont{Jackiewicz et~al.}(2003)\citenamefont{Jackiewicz,
  Blagoev, and Bedell}}]{jackiewicz2003}
\bibinfo{author}{\bibfnamefont{J.}~\bibnamefont{Jackiewicz}},
  \bibinfo{author}{\bibfnamefont{K.~B.} \bibnamefont{Blagoev}},
  \bibnamefont{and} \bibinfo{author}{\bibfnamefont{K.~S.}
  \bibnamefont{Bedell}}, \bibinfo{journal}{Phil. Mag. Lett.}
  \textbf{\bibinfo{volume}{78}}, \bibinfo{pages}{169} (\bibinfo{year}{2003}).

\bibitem[{\citenamefont{Dahal et~al.}(2005)\citenamefont{Dahal, Jackiewicz, and
  Bedell}}]{dahal2005}
\bibinfo{author}{\bibfnamefont{H.~P.} \bibnamefont{Dahal}},
  \bibinfo{author}{\bibfnamefont{J.}~\bibnamefont{Jackiewicz}},
  \bibnamefont{and} \bibinfo{author}{\bibfnamefont{K.~S.}
  \bibnamefont{Bedell}}, \bibinfo{journal}{Phys. Rev. B}
  \textbf{\bibinfo{volume}{71}}, \bibinfo{pages}{184518}
  (\bibinfo{year}{2005}).

\bibitem[{\citenamefont{Nevidomskyy}(2004)}]{andriy2004}
\bibinfo{author}{\bibfnamefont{A.~H.} \bibnamefont{Nevidomskyy}},
  \bibinfo{journal}{cond-mat/0412247}  (\bibinfo{year}{2004}).

\bibitem[{\citenamefont{Tinkham}(1996)}]{tinkham}
\bibinfo{author}{\bibfnamefont{M.}~\bibnamefont{Tinkham}},
  \emph{\bibinfo{title}{Introduction to Superconductivity}}
  (\bibinfo{publisher}{Mcgraw-Hill}, \bibinfo{year}{1996}).

\bibitem[{\citenamefont{Kotegawa et~al.}(2005)\citenamefont{Kotegawa, Harada,
  Kawasaki, Kawasaki, Kitaoka, Haga, Yamamoto, Onuki, Itoh, Haller
  et~al.}}]{kotegawa2004}
\bibinfo{author}{\bibfnamefont{H.}~\bibnamefont{Kotegawa}},
  \bibinfo{author}{\bibfnamefont{A.}~\bibnamefont{Harada}},
  \bibinfo{author}{\bibfnamefont{S.}~\bibnamefont{Kawasaki}},
  \bibinfo{author}{\bibfnamefont{Y.}~\bibnamefont{Kawasaki}},
  \bibinfo{author}{\bibfnamefont{Y.}~\bibnamefont{Kitaoka}},
  \bibinfo{author}{\bibfnamefont{Y.}~\bibnamefont{Haga}},
  \bibinfo{author}{\bibfnamefont{E.}~\bibnamefont{Yamamoto}},
  \bibinfo{author}{\bibfnamefont{Y.}~\bibnamefont{Onuki}},
  \bibinfo{author}{\bibfnamefont{K.}~\bibnamefont{Itoh}},
  \bibinfo{author}{\bibfnamefont{E.}~\bibnamefont{Haller}},
  \bibnamefont{et~al.}, \bibinfo{journal}{J. Phys. Soc. Jpn.}
  \textbf{\bibinfo{volume}{74}}, \bibinfo{pages}{705} (\bibinfo{year}{2005}).

\end{thebibliography}
\end{document}